\pdfoutput=1
\documentclass[letterpaper, 10 pt, conference]{ieeeconf}  

\IEEEoverridecommandlockouts                              

\usepackage{graphics} 
\usepackage{graphicx}
\usepackage{mathptmx} 
\usepackage{times} 
\usepackage{array}
\usepackage{amsmath} 
\usepackage{amssymb}  
\usepackage{array}
\title{\LARGE \bf
Human-Planned Robotic Grasp Ranges: Capture and Validation
}

\author{Brendon John$^{1}$, Jackson Carter$^{2}$, Javier Ruiz$^{3}$, Sai Krishna Allani$^{2}$, Saurabh Dixit$^{2}$, \\Cindy M. Grimm$^{2}$, and Ravi Balasubramanian$^{2}$
\thanks{*Supported in part by NSF Grant REU Site: Robots in the Real World, CNS 1359480}
\thanks{$^{1}${Rochester Institute of Technology, \tt\small bmj8778@rit.edu}}%
\thanks{$^{2}$Oregon State University, {\tt\small allanis, carterj, dixits, Cindy.Grimm, Ravi.Balasubramanian, @OregonState.edu}}%
\thanks{$^{3}$UC Santa Cruz {\tt\small jamaruiz@ucsc.edu}}%
}

\begin{document}

\maketitle
\thispagestyle{empty}
\pagestyle{empty}

\begin{abstract}
Leveraging human grasping skills to teach a robot to perform a manipulation task is appealing, but there are several limitations to this approach: time-inefficient data capture procedures, limited generalization of the data to other grasps and objects, and inability to use that data to learn more about how humans perform and evaluate grasps.  This paper presents a data capture protocol that partially addresses these deficiencies by asking participants to specify {\em ranges} over which a grasp is valid.   The protocol is verified both qualitatively through online survey questions (where 95.38\% of within-range grasps are identified correctly with the nearest extreme grasp) and quantitatively by showing that there is small variation in grasps ranges from different participants as measured by joint angles, contact points, and position.  We demonstrate that these grasp ranges are valid through testing on a physical robot (93.75\% of grasps interpolated from grasp ranges are successful).
\end{abstract}

\section{INTRODUCTION}
Humans are excellent at physical manipulation, robots less so.  Thus, it is appealing to use humans to ``teach'' robots how to manipulate objects.  This difference in abilities is not just a result of the human and the robot using  different manipulators; that is, the human hand versus a robot hand.  There is a difference in the strategies humans use when compared with the strategies used by current robotic grasp planning.  For example, the work in~\cite{Balasubramanian-TRO-2012} showed that humans use even robotic manipulators better than current automatic algorithms for unstructured grasping tasks.  Unfortunately, it is unclear what strategies humans use to choose a grasp.  Furthermore, elucidating those strategies is challenging.  Thus, most methods used to train robots rely on some form of human demonstration of {\em specific grasp examples for specific tasks} using a physical robot~\cite{ekvall2004interactive}.  Unfortunately, this method of collecting human input only results in a small number of specific grasps for specific objects.  It is not clear how to generalize this data or make it robust to small perturbations in robot posture or object location. 

In this paper, we outline a novel protocol for collecting human-generated grasp examples that capture not just a single grasp, but an acceptable {\em variation} of that grasp from many participants.  This protocol results in a more time-efficient  capture of example grasps and data that is easier to generalize.  Specifically, the protocol: 1) Captures {\em regions} of good grasps {\em and} bad grasps, instead of single ("optimal" or first-choice) grasp instances.  2) Captures additional human-centric information to elucidate: {\em how} the participant arrived at a specific grasp, how that grasp compares to a grasp that the participant would do with their own hand, and what the participant looked at when performing the grasping task. 3) Supports capture of manipulation tasks where one robot hand can only perform part of the manipulation, such as picking a small object off the table or re-positioning a grip.  Post-protocol, we also use a survey-based approach for obtaining information on the similarity of grasps collected in order to better group them.

Capturing a grasp {\em region} instead of a single grasp, and capturing many bad grasps, is useful for two reasons.  First, sophisticated machine learning algorithms can use this data to better define regions of valid inference, such as the {\em boundaries} of where a successful grasp will fail.  Increasing the number of negative training examples is also important to improve machine learning results (for example, where not to focus grasp search).  Second, this provides insight into how humans group grasps.  This information can be used in human-robot interaction applications to search for grasp metrics that match those groupings~\cite{billard2008robot}. 

Identification of similar grasps across people and objects makes it easier to generate robust grasps by providing a partitioning of the data set into grasps where simple linear interpolation makes sense.  In our experiments, the first few grasps the participants provided tended to be similar to the ones given by other participants, providing evidence that these grasps are reliable in the view of the participants.  


%
%

We use a mix of a think-aloud protocol, prompting questions, and eye-tracking to capture human thought processes and visual attention during the grasping tasks.  We also ask the participants to perform the grasp with their own hand as well as with the robotic hand, and explain the differences. This captures both high-level cognitive processes (what the participants {\em think} they're doing) and low-level actions (actual physical and perceptual actions). Although we do not discuss analysis of this data in detail in this paper, preliminary analysis shows that this information is useful for identifying good views of objects and an initial (partial) mapping of existing grasp metrics to human considerations.~\cite{matt2016paper}
%
%
%

The proposed protocol supports capturing a variety of grasping and manipulation tasks, such as picking an object up from the table, receiving the object from another hand, and manipulating the object.  To test our protocol, we chose one structured task for each object (pick the object up from the table and place it on a box), and one more natural task suitable for that object (such as pouring for the water pitcher). The latter was deliberately free-form --- the participants were free to place the object in the robot hand or pick it up from the table. To support this more free-form interaction we used a Kinect\texttrademark{} sensor to track the object's location relative to the hand (in addition to tracking the joint angles of the robot hand and arm). 

To validate the grasp ranges collected, we focus on the following questions: 1) How do we identify and verify similarity of grasps across participants?  2) What is the variation in the range of grasps that participants prefer for an object? 3) How do grasps sampled from the participants's grasp ranges perform in practice? 
%
%
%

%
%

\section{RELATED WORK}
There has been significant progress in the domain of robotic grasping and manipulation both in terms of hardware~\cite{Dollar-Howe-2010, Birglen-Book, brown2010universal} and software development~\cite{Saxena-Driemeyer-Ng-IJRR-2008, lopez2005grasp, leon2010opengrasp, chitta2012moveit}.  However, prior work has shown that even in a laboratory environment with almost perfect information for grasp planning, robotic grasping performance only succeeds about 75\% of the time; that is, one in four grasps fail~\cite{Balasubramanian-TRO-2012}.  The primary reason for this poor performance is that robot grasps are not robust enough; that is, small differences in object shape or object position cause the object to, say, slip out during the grasping process.  There has been significant effort to address these issues using physics-based heuristics and brute-force search algorithms to find more robust grasps~\cite{bohg2013data, cornelia2005determining}.   However, a big hurdle that these methods face is that grasping tasks exist in a continuous space with significant uncertainty (for example, arising from unknown friction and compliance in the gripper fingertips) and noise (for example, arising from the actuator position errors or warped sensory perception).   Thus, even if perfect discrete grasps are created, a millimeter or two of positioning error will cause the grasp to fail.  This is precisely the reason why this work collects grasp {\em ranges} from humans and identifies grasps that participants rank as similar.  This provides evidence that the grasp is robust to small variations and grasps do not live on discrete islands.

Prior work has also explored ``learning from  demonstration (LfD)'', where humans teach robots to advance robot performance~\cite{RobotLearnGrasp2004,Argall2009469,bluethmann2004building,niekum2012learning}.   The key idea is to identify generalized control policies based on  specific examples.  However, these approaches for gathering data are time-inefficient and require vast amounts of data to learn policies~\cite{5509855}.   The approach taken in this work of collecting grasp ranges fits well with existing LfD techniques~\cite{suay2012practical}, since it taps into human intuition for generalizing grasps around optimal grasps.
%
%


\begin{table}
  \caption{Data capture methods and usage}
  \label{tab:data}
\begin{tabular}[c]{| m{1.6cm} | m{5.8cm} |}    \hline
\multicolumn{1}{|>{\centering\arraybackslash}m{1.6cm}|}{\textbf{Data}} & \multicolumn{1}{|>{\centering\arraybackslash}m{5.8cm}|}{\textbf{Usage}} \\ \hline 
    Gaze Data  & Identify which regions of the scene are fixated on most while grasping.  \\ \hline 
    Eye Tracker Scene Video  & Used to get perspective view of subject, provides snapshots of specified grasps. \\ \hline 
    Kinect Depth Data  & Object tracking and grasp reconstruction. \\ \hline 
    Barrett Hand and Arm Data  & Joint angles and positions used for recreating grasps performed in the study. \\ \hline 
    Human Specified Ranges  & Ranges can be utilized to improve machine learning models, and tested using an automated system. \\ \hline
    Good and Bad Trials & Bad grasp trials provides samples of grasps to avoid when working with models or simulations.\\ \hline
  \end{tabular}
\end{table}











\section{STUDY PROTOCOL AND METHODOLOGY}
In this section, we describe the human-subject experiment protocol in terms of the tasks, phases, specific queries, data captured,  protocol management, and participants.   At a high level, participants were asked to perform a specific manipulation task with a given object. They accomplished this task both with their own hand and with the robotic hand, but we used only the data from the trials with the robot hand.  The participants were asked to show as many ``good'' and ``bad'' grasps as they could.  In addition to performing the manipulation, they were asked to specify the valid range for each unique grasp (if such a range existed).  We captured four data streams: the participant's eye gaze, audio, arm and hand posture, and a 4D video (color plus depth) (see Table~\ref{tab:data}), but use only the robot arm and hand position and 4D video in this paper.

\subsection{Objects and tasks}
\begin{figure}
  \centering
  \includegraphics[width=0.9\linewidth]{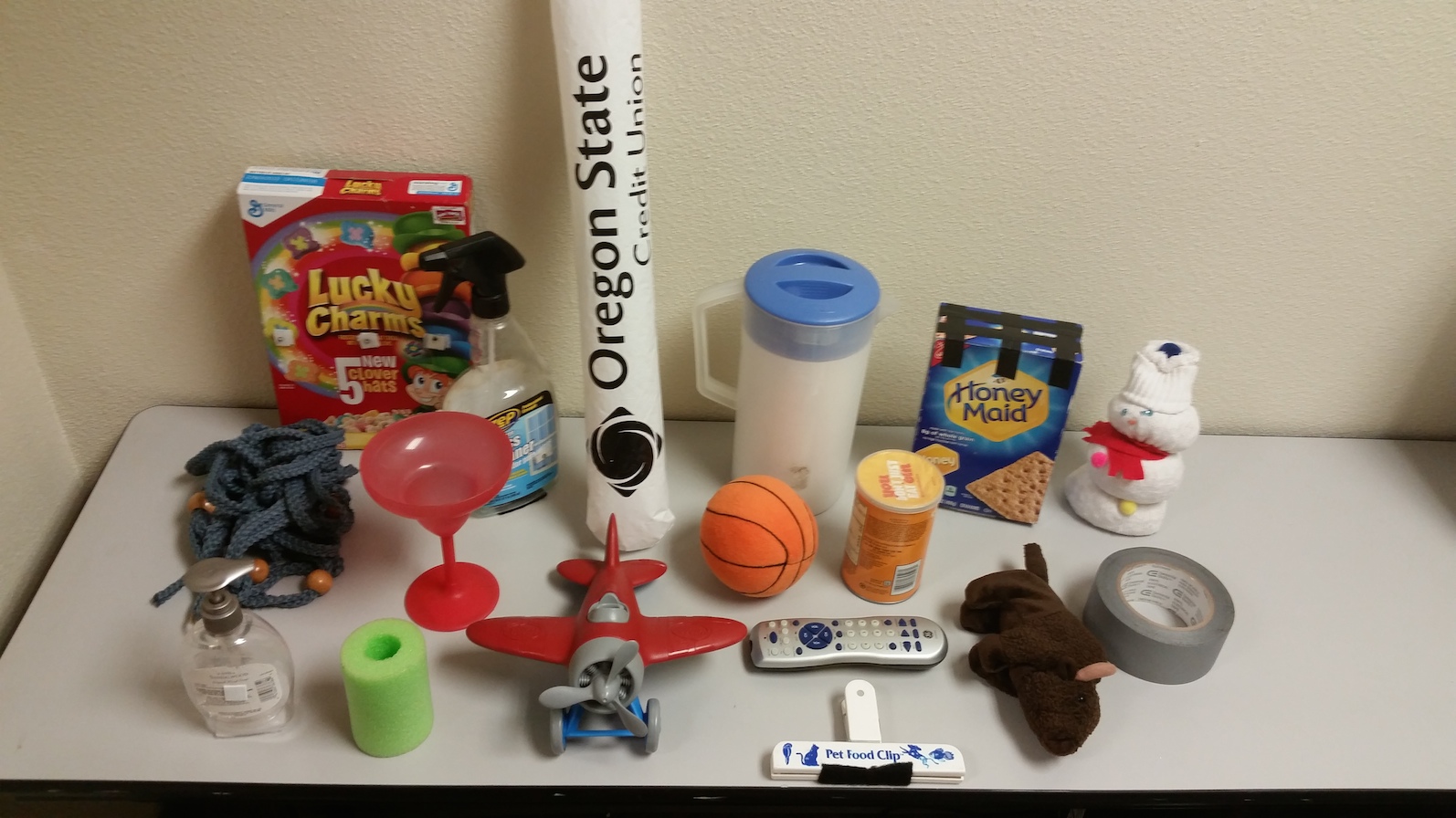}
  \caption{Objects used for the study.}
  \label{fig:objects}
\end{figure}

The objects used in this protocol are shown in Figure~\ref{fig:objects}.  For each object, the participants were asked to perform two tasks. The first was to pick up the object from the table and place it on a nearby box. The second task was object-specific (see Table~\ref{tab:objects}). These object-specific tasks are tasks or actions that are commonly associated with each object, such as throwing a ball, squeezing a trigger, or handing the object to someone.

For some of the tasks (such as picking up the snowman from the table) the robotic hand was not physically capable of performing the task with the object lying on the table.   In this case, the participants were allowed to pick up the object and put it in the robot's hand.  However, the participants were required to actually perform the task with the robotic arm, such as grasping the object with the robot hand and then moving the arm to perform the task.  This ensured that the given grasps were successful, even if placing the object into the grasps required assistance. \\
  \vspace{-0.1in}

\begin{figure*}
  \centering
  	\includegraphics[width=0.125\textwidth]{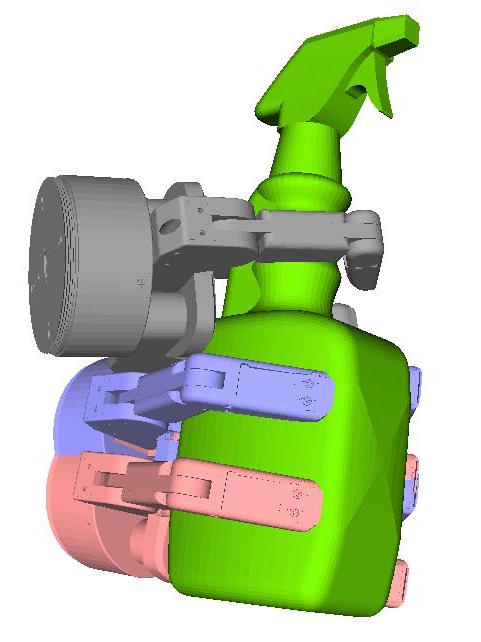}
  	\includegraphics[width=0.125\textwidth]{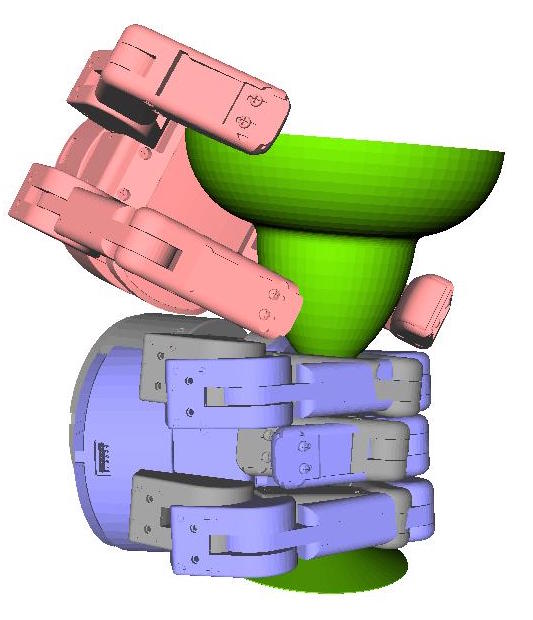}
    \includegraphics[width=0.125\textwidth]{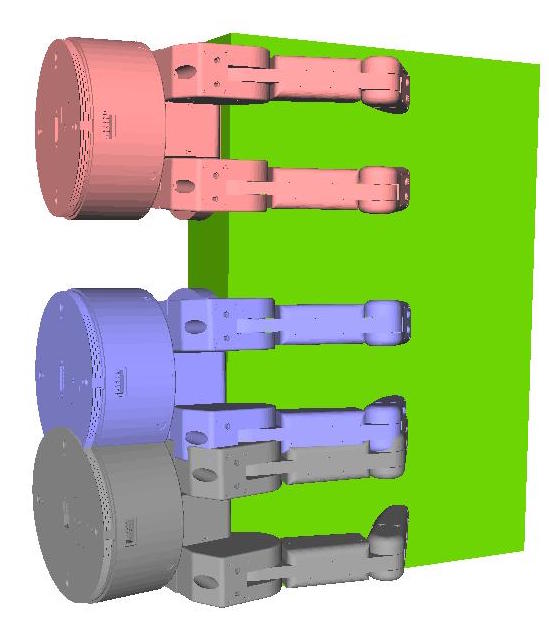}
    \includegraphics[width=0.125\textwidth]{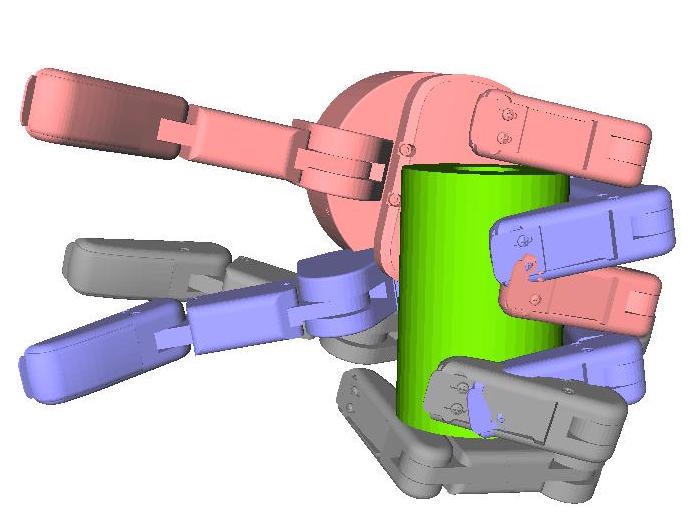}
    \includegraphics[width=0.125\textwidth]{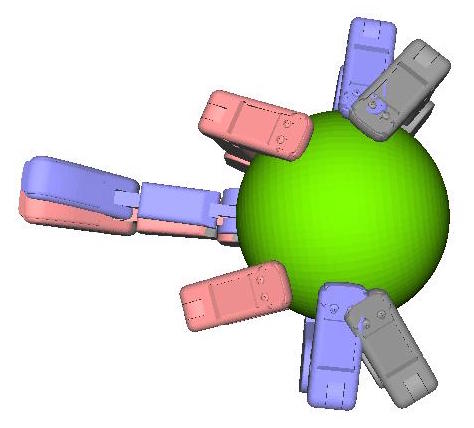}

  \caption{1-5: Five objects with the grasp range given by one participant. The violet hand is the original grasp, the gray and red show the first and second extreme grasp locations.}
  \label{fig:GraspRangesFigure}
  \vspace{-0.1in}
\end{figure*}


\begin{table}
\centering
  \caption{Object-specific tasks and number of grasps captured (including pick-up task).}
    \label{tab:objects}
  \begin{tabular}{| m{2.2cm} | m{3.4cm} | m{0.6cm} | m{0.6cm} |}
\hline
\multicolumn{1}{|>{\centering\arraybackslash}m{2.2cm}|}{\textbf{Objects}} & \multicolumn{1}{|>{\centering\arraybackslash}m{3.4cm}|}{\textbf{Natural Task}} & \multicolumn{2}{|>{\centering\arraybackslash}m{1.2cm}|}{\textbf{Grasps}} \\  
\multicolumn{1}{|>{\centering\arraybackslash}m{2.2cm}|}{} & \multicolumn{1}{|>{\centering\arraybackslash}m{3.4cm}|}{\textbf{}} & \multicolumn{1}{|>{\centering\arraybackslash}m{0.6cm}|}{\textbf{Good}}& \multicolumn{1}{|>{\centering\arraybackslash}m{0.6cm}|}{\textbf{Bad}} \\  
\hline \hline
    Water Pitcher  & Pour water out of pitcher & 11 & 1 \\ \hline
    Spray Bottle  & Pull trigger to spray & 14 & 22 \\ \hline
    Margarita Glass  & Drink out of glass & 14 & 14 \\ \hline
    Cereal Box & Pour cereal out of box & 12 & 18 \\ \hline
    Cracker Box & Pour crackers out of box & 15 & 7 \\ \hline
    Television Remote & Press power button on remote & 11 & 14 \\ \hline
    Toy Plane & Pretend to fly plane around & 13 & 19 \\ \hline
%
%
%
%
    Food Clip & Open clip (to close bag) & 10 & 3 \\ \hline
    Soap Dispenser  & Press down on nozzle to dispense soap & 10 & 5 \\ \hline
    Foam Cylinder & Throw object overhand & 16 & 15 \\ \hline
    Bison Plush Toy* & Hand toy to someone & 5 & 0 \\ \hline
    Plush Ball  & Throw ball underhand & 19 & 10 \\ \hline
    Thunder Stick* & Hit something with it & 10 & 0 \\ \hline
    Sock Doll & Hand doll to someone & 16 & 15 \\ \hline
    Decorative Cord & Hang cord by its metal ring & 5 & 6 \\ \hline
    Tape Roll & Support tape roll for ripping tape off & 11 & 4 \\ \hline
    {\bf Total} & & 192 & 153\\ 
    {\bf Mean} & & 11.29 & 9\\ \hline
  \end{tabular}
\begin{flushleft}\textnormal{Note: Objects with * do not have any bad grasps}\end{flushleft}
\end{table}
  \vspace{-0.1in}

\subsection{Phases}
Our study protocol is designed to capture both human grasping and human-planned robotic grasping. To do this, the study features a training phase and two distinct capture phases.

In the training phase, which preceded data capture, participants were asked to familiarize themselves with the robot hand by moving it around and adjusting the fingers, specifically showing them how to change the spread of the fingers. Although the arm was gravity compensated, it did not always maintain its position when the hand was opened and closed due to the hand's shifting center of mass, so participants were also given instructions to ask for help in supporting the hand if needed. 
%
%
%
%

In the first capture phase, the participants use their own hands to grab an object, while in the second, the participant physically positions the robotic arm and hand to grasp the object. The order of the two phases was randomized for each object. 

For the human-hand grasping phase, participants were asked to use only their thumb and first two fingers to mimic the three fingers of the robotic hand.  This paper does not the human-hand grasping data, which will be analyzed in future work.

The robotic grasping phase is further split into three phases: Specifying a pick-up grasp, performing the manipulation task, and specifying ``bad'' grasps. For every grasp we also asked the participants to specify a range for that grasp  (if the participant felt there was one) before moving on. 
%
%
%
%

\noindent {\bf Grasp ranges:} For each specified grasp (both good and bad), the participant was also asked to provide a range along the object for which that particular grasp could still be applied (e.g. along the side of a box --- see Figure~\ref{fig:GraspRangesFigure})) . Participants were also asked if there were any rotational or symmetric ranges for their grasps.  In the good grasp phase, these ranges are the areas on the object that the participant believes the grasp would still be successful. In the bad grasp phase this range represents the region that a particular grasp would still be ineffective. Not all grasps had ranges, for example, holding the spray bottle to spray.

\noindent {\bf Bad grasps:} Participants were asked to ``teach'' the robot how {\em not} to pick up the object by demonstrating bad grasps. They were allowed to specify as many grasps as they wanted, but were asked to demonstrate the grasp failing for each one\footnote{Some participants accidentally found good grasps this way.}. 

Because of participant time constraints and the eventual fatigue induced by the procedure, not all subjects were able to complete all phases of the study. Occasionally, participants would complete the good grasp phase of the study, but not have time to complete the bad grasp phase. Thus, some objects do not yet have bad grasp data and associated statistics.

\subsection{Prompts and think-aloud}

The subjects were asked to think out loud as they performed the study to provide insight into what they were thinking of while performing the grasping tasks.
%
%
For the pick up the object task, the participants were asked to move the object using the robotic hand after finalizing the grasp. For the other tasks, they were not required to perform the task, but simply needed to position the hand. They were given explicit permission to pick up the object, position it how they wanted, and to use their other hand if they needed two hands. Participants were always given the option of specifying another grasp for that task (if they could think of one). 
%
%
%
%

We had one further prompt, asked at the end of every good grasp: ``Is this grasp exactly what you wanted? Or are the finger placements slightly different that what you were intending? (How so?)''. This prompt is aimed at disambiguating how much the robotic hand limitations affected the participant's grasp choice.

\subsection{Data capture: equipment and procedure}

\begin{figure}
  \centering
  
  \includegraphics[width=0.8\linewidth]{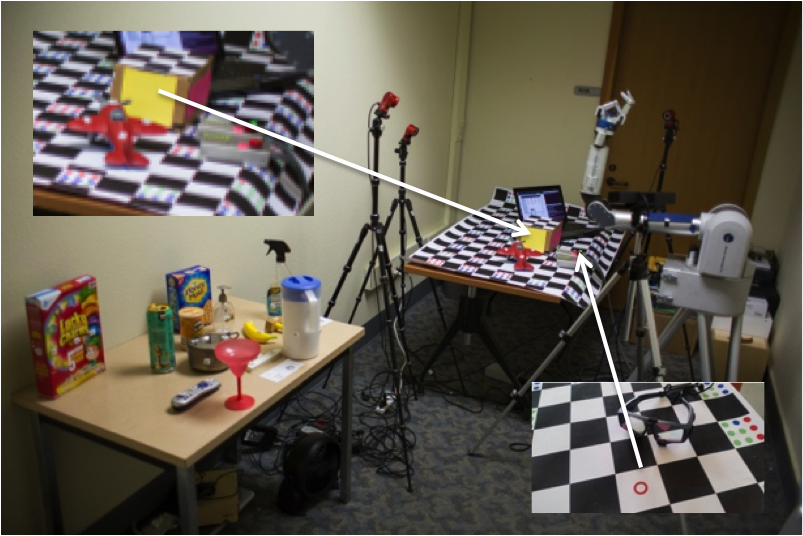}
  \caption{Study set up. The table included a checkerboard pattern for further calibration. The red circle was used to calibrate the eye tracker. The box on the table was used in the object placement tasks. Participants were initially seated on the opposite side of the table from the Kinect sensor; they were not required to remain seated.}
  \label{fig:setup}
  \vspace{-0.1in}
\end{figure}
The equipment used for this study included a pair of SMI Eye Tracking Glasses 2.0 to collect eye-gaze data, a Microsoft Kinect v2 to collect depth data, and a Barrett Whole Arm Manipulator (WAM) with BH280 BarrettHand to perform the robotic grasping.  We also included spatial calibration patterns for the table and box (see Figure~\ref{fig:setup})  to ensure calibration between video data sources. Data captured, and its uses, are summarized in Table~\ref{tab:data}.

%
%
%
%
%
%
%

\noindent {\bf Eye tracking:} The SMI glasses record both where the user is looking and what they are looking at. The data is recorded as a 960x720 video stream at 30 Hz, plus an eye gaze location for each video frame (as $x,y$ image coordinates). The eye gaze data also includes other information such as pupil diameter, fixations, and saccades.
%
%
%
%
The eye tracker has to be fit to the person's head (similar to goggles) using two nose pieces and calibrated to their eyes. To perform the calibration the participant was asked to sit down in front of the table and fixate on a red dot on the table (see Figure~\ref{fig:setup}). This one point calibration was performed using the SMI software. We checked the calibration at the end of each grasp trial by having the participant focus on the red dot again.  The eye-gaze tracking data is not used in this paper.
%
%
%
%

\noindent {\bf Arm and hand tracking:}
We used a Barrett WAM and BarrettHand (BH-280) in the study. The arm is backdrivable and gravity compensated; that is, the arm location can be physically adjusted with ease.  However, the BarrettHand's fingers cannot be physically adjusted from external forces (only through its motors).  We used a physical set of three sliders to control how much each finger was closed, and a knob to control the spread of the fingers. Note that the two joints of the finger are controlled with one actuator.
%
%
%
%

The Barrett WAM itself features an onboard Linux PC which both compensates for gravitational effects on the arm linkage and continuously streams joint positions to the main Linux PC via ethernet. The onboard PC includes a timestamp with each joint position measurement, so care was taken to synchronize the onboard PC's clock with the main Linux PC's clock using the program chrony.

The gravity compensation function was calibrated with the hand open to reduce the need to support the arm when the hand changed configuration; however, it was still sometimes necessary to externally stabilize the hand when adjusting the fingers. We provided explicit instructions to the participants to ask one of the study team members to help if this happened.


\noindent {\bf Kinect 3D scene data}
The Microsoft Kinect provides a $960x540$ RGB image with depth data for each pixel at 30fps. The depth is from time-of-flight, so is (approximately) correct in real-world terms. The depth data, along with an estimate for the camera's orientation with respect to the scene, was used to construct a 3D point cloud of the scene. We use the point cloud to find the object's location with respect to the hand (see Section~\ref{sec:PCArmCalibration}).

\noindent {\bf Audio:} The eye-tracker records audio with the video. 


\noindent {\bf Temporal data stream alignment:} 
The Robot Operating System (ROS) was used to integrate and store data feeds from the Barrett WAM, the BarrettHand, and the Kinect. RGB and depth images are recorded continuously during the grasping phases along with the robot's arm and hand joint positions when it is being manipulated. When participants finalize a grasp, a fixed-format annotation is (manually) added to the data stream along with a timestamp for future synchronization. Additionally, a more general text annotation system is used to flag grasps in order to make them easier to find in the video data.  The eye-tracking system was temporally synced with the other data streams using a high-pitched beep.

%
%


\subsection{Protocol management and flow}
The study is designed to be run by two researchers; one for Eye tracker and other for Ubuntu ROS PC.

The average time for a data collection session was an hour and a half, covering two grasps each for three or four objects. The maximum time was capped at two hours due to eye strain generated by the eye tracking glasses, as well as general fatigue from performing the experiment. 
%
%
%
%
%
%
Although we limited a single session to two hours, several participants did two sessions. The single-session participants were asked to perform good grasps on two objects first, and then bad grasps with two objects. For two-session participants we did good grasps for as many objects as possible in the first session, and bad grasps in the second.

\begin{figure}
  \centering
  \includegraphics[width=0.9\linewidth]{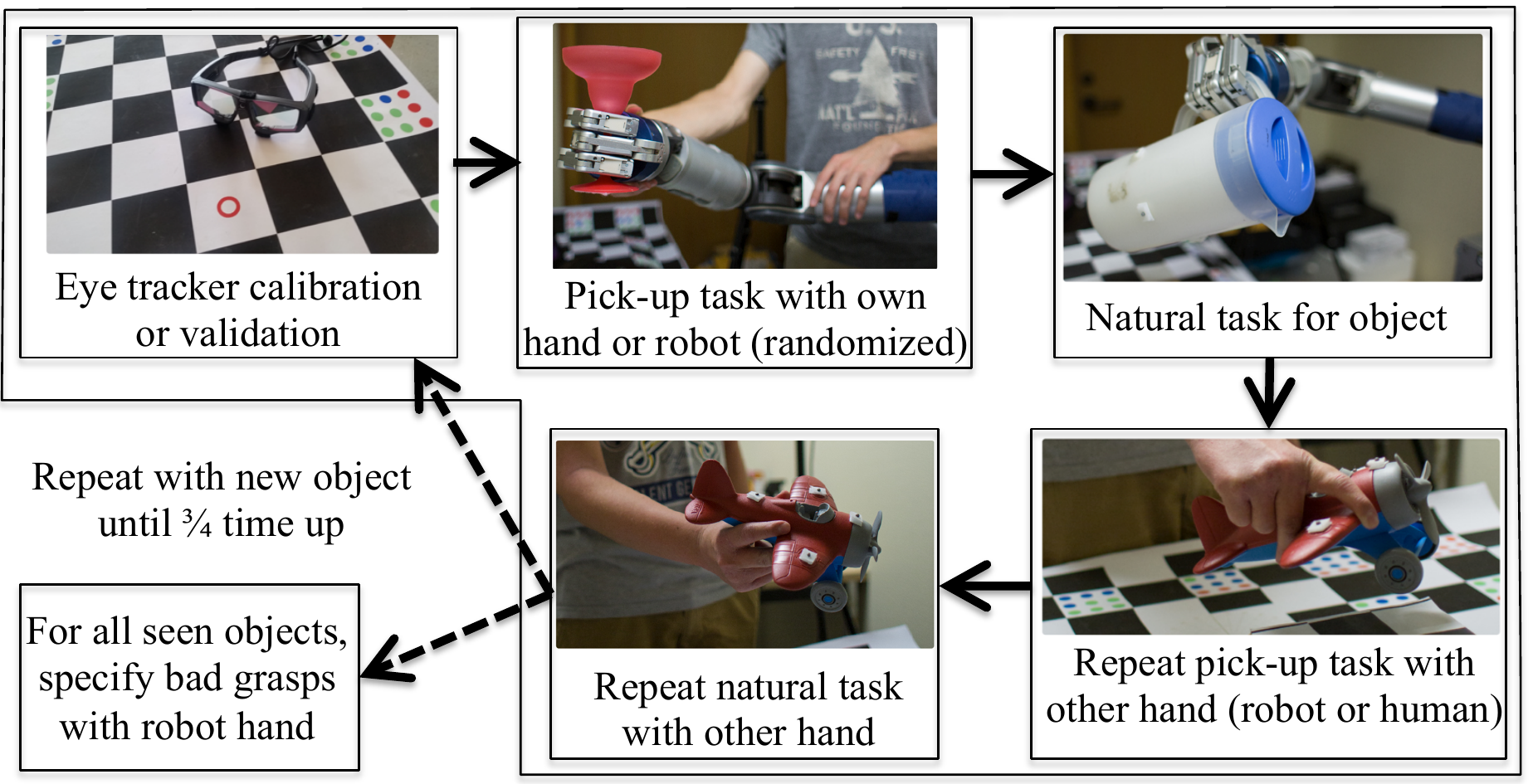}
  \caption{Flow chart of study procedure.}
  \label{fig:procedure}
\end{figure}

The general flow of the study can be seen in Figure~\ref{fig:procedure} and is also outlined in the list below.

\begin{enumerate}
\item Participant enters room and signs consent form.
\item Brief training session with a test object.
\item Eye tracking calibration performed.
%
%
\item Study trials explained to participant.
\begin{enumerate}
\item Object placed on table, and participant told to use robot hand (group 1) or their own hand (group 2) to perform pick up task.
\begin{enumerate}
\item Pick up task performed. 
\item Grasp range specified
\item Repeat until no new grasps.
\end{enumerate}
\item Natural task explained
\begin{enumerate}
\item Natural task performed. 
\item Grasp range specified
\item Repeat until no new grasps.
\end{enumerate}
\item Object-tasks (a-b) repeated with human hand (group 1) or robot hand (group 2)
\item Eye tracking recording stopped, and re-calibration if needed. 
\end{enumerate}
\item Repeat a-d with as many objects as possible (approximately 1.5 hours)
\item Bad grasp prompt
\begin{enumerate}
\item Participant given first object from good grasp trials
\begin{enumerate}
\item Bad grasp and range specified.
\item Repeat until no new grasps.
\end{enumerate}
\item Repeat a) with next object from good grasp trials
\end{enumerate}
\item Eye tracking recording stopped, all other data collection ended.
\end{enumerate}

\subsection{Participants}
We recruited 13 participants, ranging in age from 16 to late 50's, all with normal or corrected to normal with contacts vision (it is not possible to simultaneously wear regular eye glasses and the Eye-gaze ones). On average participants specified 2-4 good grasps per object, with a high of 8, and 3-5 bad grasps (see Table~\ref{tab:participants}). The largest number of grasps any participant specified for an object was 16 for the Foam Cylinder. The fewest number of grasps specified for an object was 4 for the Margarita Glass. Of all 294 grasps collected, 179 had extremes (therefore comprising a grasp range) while 115 did not.


\begin{table*}
\begin{center}
\caption{Average number of each type of grasp given per participant per object.}
  \label{tab:participants}
\begin{tabular}{| l | ccccccccccccc | l |}    \hline
Type & \multicolumn{12}{c}{Average number of grasps per participant} & & Mean \\ \hline
Good & 4 & 6.25 & 3.25 & 4.5 & 4.8 & 7 & 3 & 3.5 & 3.75 & 2.5 & 2.5 & 5.67 & 8 & 4.5 \\
Bad & 0 & 7.5 & 4 & 5.75 & 3.2 & 0 & 3.75 & 4.75 & 4.5 & 1.5 & 2 & 2 & 0 & 3.0 \\ \hline
  \end{tabular}
  \vspace{-0.1in}
  \end{center}
\end{table*}

\section{VERIFICATION OF PROTOCOL}
In this section, we provide information on how we verified initial measures of grasp similarity across participants, how we verify and use the grasp ranges, and how we verify using physical shake tests and participant evaluation the quality of grasps generated from the grasp ranges.

Data collection of this sort is very time consuming.   Furthermore, there is an inherent trade-off between data reliability and data variability.  Specifically, we can collect a lot of (possibly redundant) data about one object manipulation task, or collect (possibly biased or missing) data for many object manipulation tasks.  We have settled on the following protocol for determining how many participants is enough.  We collect data from a minimum of two participants for each object. If the participants both specify  one (or more) ``similar'' grasps (see Section~\ref{sec:consistency}) then we mark that object as complete and move to the next one. If the participants gave different grasps, then we continue until we see no new grasps. In practice, this did not happen.

The motivation behind this approach is as follows: 1) In our formative studies, participants were remarkably consistent in producing the same grasp and grasp range.  While some participants were more creative (specifying many grasps), their first grasp was almost always the same as other participants. 2) Since we are explicitly collecting a {\em range} for each grasp, we do not need a lot of individual grasp examples.  This range can then be verified with other, less time-intensive methods such as using crowd sourcing (showing images of example grasps within that range) or semi-automated physical shake testing~\cite{Balasubramanian-TRO-2012}. 
%
%
%


%
%
%
%
%


\subsection{Grasp similarity}
\label{sec:consistency}
We evaluate the similarity of grasps in three stages: 1)~An initial qualitative assessment by the experimenters of the grasps provided for five objects to determine if the claims that participants provide similar grasps is true.  2)~A subsequent quantitative assessment using joint angles, contact points, and positions. 3)~Shake test of grasps grouped by similarity. 

\begin{figure*}
  \centering
\begin{minipage}[b]{0.16\textwidth}
  \includegraphics[width=\textwidth]{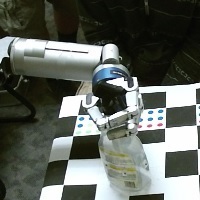}
\end{minipage}
\begin{minipage}[b]{0.16\textwidth}
  \includegraphics[width=\textwidth]{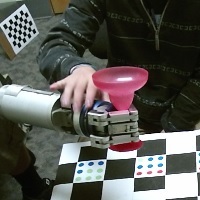}
\end{minipage}
\begin{minipage}[b]{0.16\textwidth}
  \includegraphics[width=\textwidth]{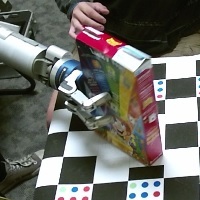}
\end{minipage}
\begin{minipage}[b]{0.16\textwidth}
  \includegraphics[width=\textwidth]{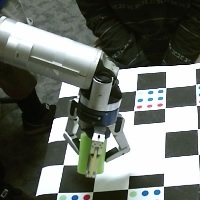}
\end{minipage}
\begin{minipage}[b]{0.16\textwidth}
  \includegraphics[width=\textwidth]{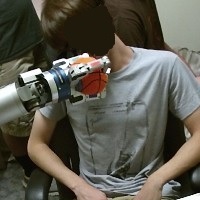}
\end{minipage}

\begin{minipage}[b]{0.16\textwidth}
  \includegraphics[width=\textwidth]{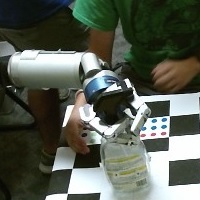}
\end{minipage}
\begin{minipage}[b]{0.16\textwidth}
  \includegraphics[width=\textwidth]{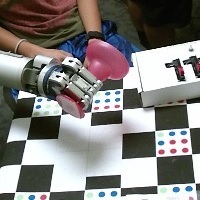}
\end{minipage}
\begin{minipage}[b]{0.16\textwidth}
  \includegraphics[width=\textwidth]{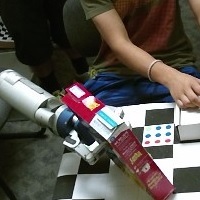}
\end{minipage}
\begin{minipage}[b]{0.16\textwidth}
  \includegraphics[width=\textwidth]{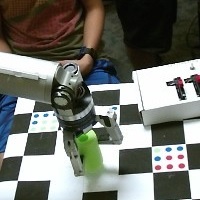}
\end{minipage}
\begin{minipage}[b]{0.16\textwidth}
  \includegraphics[width=\textwidth]{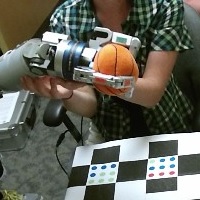}
\end{minipage}

\begin{minipage}[b]{0.16\textwidth}
  \includegraphics[width=\textwidth]{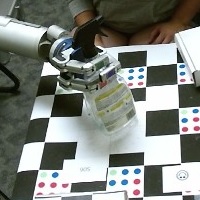}
  \begin{center}
  \textnormal{\footnotesize{(a) Spray Bottle}}
  \end{center}
\end{minipage}
\begin{minipage}[b]{0.16\textwidth}
  \includegraphics[width=\textwidth]{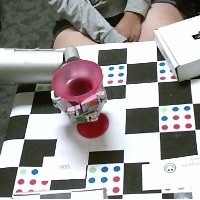}
  \begin{center}
  \textnormal{\footnotesize{(b) Margarita Glass}}
  \end{center}
\end{minipage}
\begin{minipage}[b]{0.16\textwidth}
  \includegraphics[width=\textwidth]{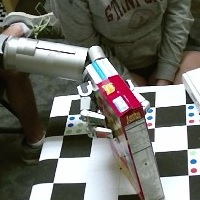}
  \begin{center}
  \textnormal{\footnotesize{(c) Cereal Box}}
  \end{center}
\end{minipage}
\begin{minipage}[b]{0.16\textwidth}
  \includegraphics[width=\textwidth]{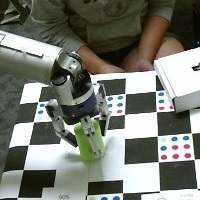}
  \begin{center}
  \textnormal{\footnotesize{(d) Chunk of Foam}}
  \end{center}
\end{minipage}
\begin{minipage}[b]{0.16\textwidth}
  \includegraphics[width=\textwidth]{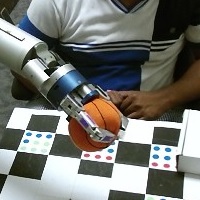}
  \begin{center}
  \textnormal{\footnotesize{(e) Ball}}
  \end{center}
\end{minipage}
  \caption{Images of similar grasps for three participants for five objects.}
  \label{fig:ValidationImage}
  \vspace{-0.1in}
\end{figure*}

Stage one used images of grasps of five of the thirteen objects (cereal box, ball, chunk of foam, spray bottle, wine glass) to qualitatively segregate  grasps into groups based on their similarity (see Figure~\ref{fig:ValidationImage}). Stage two is quantitative: Although we do not have enough participants per object to make any statistically significant claims, these numbers form a base-line for quantifying the qualitative notion of similar grasps.  All quantitative measures are normalized (zero is no variation, one is maximum variation). See Table~\ref{tab:Validation}. In stage three we performed shake tests of groups of similar grasps.
%
We performed similarity analysis based on following metrics:
%
%
%

\noindent {\bf Joint angle:} We measure the difference to the mean over the available range of the joint, averaged over all of the joints for the hand (finger spread, finger joint angles). 

\noindent {\bf Contact points:} We determine contact points at the fingertip, pad, and finger joints using OpenRAVE~\cite{Diankov_2008_6117}, specifically, we calculate intersections with the object and the hand after alignment (Section~\ref{sec:PCArmCalibration}). The measure we used was the average of all-pairs edit distance (number of changed contacts over number of contacts).

\noindent {\bf Positions: } For this measure, we aligned the hand to the coordinate system of the object.  For objects with symmetry we used the palm's location to resolve the ambiguity (eg, the vector to the palm to align the glass' horizontal axis).  We chose 14 positions (points) on the hand (three along each finger, five on the palm) and calculated their average distance from the mean, divided by the bounding box of the object.

\begin{table}
\centering
\caption{Standard deviation based numerical evaluation of similarity of grasps (three participants each).}
\begin{tabular}{|l|cl|ccc|}
\hline
\multicolumn{1}{|>{\centering\arraybackslash}c|}{Object} & Grasp & Type & Joint  & Contact  & Positional \\
 &  set &  & angle &  points & distance \\
\hline
Spray & 1 & optimal & 0.10 & 0.07 & 0.12 \\
 & 2 & optimal & 0.08 & 0.00 & 0.03 \\
Glass & 1 & optimal & 0.08 & 0.13 & 0.12 \\
 & 2 & optimal & 0.07 & 0.04 & 0.16 \\
 & 2 & extreme1 & 0.06 & 0.03 & 0.14 \\
 & 3 & optimal & 0.10 & 0.15 & 0.21 \\
Box & 1 & optimal & 0.06 & 0.00 & 0.03 \\
 & 1 & extreme1 & 0.12 & 0.00 & 0.02 \\
 & 2 & optimal & 0.06 & 0.03 & 0.16 \\
 & 2 & extreme1 & 0.06 & 0.00 & 0.23 \\
Foam & 1 & optimal & 0.08 & 0.00 & 0.34 \\
 & 1 & extreme1 & 0.09 & 0.02 & 0.38 \\
 & 1 & extreme2 & 0.09 & 0.00 & 0.36 \\
 & 2 & optimal & 0.20 & 0.18 & 0.35 \\
 & 2 & extreme1 & 0.04 & 0.12 & 0.52 \\
 & 2 & extreme2 & 0.04 & 0.10 & 0.20 \\
 & 3 & optimal & 0.03 & 0.00 & 0.05 \\
Ball & 1 & optimal & 0.05 & 0.09 & 0.20 \\
 & 1 & extreme1 & 0.08 & 0.10 & 0.29 \\
 & 1 & extreme2 & 0.10 & 0.07 & 0.22 \\
 & 2 & optimal & 0.04 & 0.09 & 0.08 \\
 & 2 & extreme1 & 0.08 & 0.13 & 0.19 \\
 & 2 & extreme2 & 0.03 & 0.00 & 0.05 \\
 & 3 & optimal & 0.03 & 0.03 & 0.12 \\
 & 3 & extreme1 & 0.24 & 0.15 & 0.26 \\
 & 4 & optimal & 0.07 & 0.10 & 0.07 \\
 & 4 & extreme1 & 0.20 & 0.12 & 0.18 \\
 \hline
 Mean & - & - & 0.08 & 0.07 &   0.19 \\
\hline
\end{tabular}
\label{tab:Validation}
\vspace{-0.1in}
\end{table}



\subsection{Grasp ranges}
We validated grasp ranges for the five objects both qualitatively (using an online survey) and quantitatively using a shake test (see following section). To generate intermediate grasps we employed the following algorithm implemented in OpenRAVE: Linearly interpolate the joint angles for the two (or more) grasps to be interpolated. Deal with any intersections by moving the palm, followed by the fingers, out of the surface. Then close the fingers until contact is made for those fingers that have contact points in one (or more) of the original grasps.
%
%
%

We validated the human-evaluated similarity of the interpolated grasps to the original grasps using a survey. Participants were shown three images: the original grasp, an interpolated grasp, and the corresponding extreme grasp. Participants were asked to pick which grasp (original or extreme) was most similar to the interpolated one. We tested two interpolated grasps; one that was 1/3 between the two and one that was 2/3. Figure~\ref{fig:surveyResults}, left, shows the results; in nearly all cases the survey participants correctly identified which grasp the interpolated one was most similar to (from a linear interpolation standpoint). 
%
%
%
%
%
%

We also tested these interpolated grasps for effectiveness using a shake test (summarized in the following section; results of the shake test are in Table~\ref{tab:shakeTest}).

\subsection{Shake tests}
We validated the original, extreme, and interpolated grasps using a shake test where the robot hand automatically picked up the object from the table and shook it (five times per test). The exception to this is the grasp with the hand under the ball; for this test we suspended the ball in the air using a string. We used OpenRAVE's inverse kinematics module to calculate the joint states for the robot arm to take the hand to the object. The robot's hand position itself was the one specified by the participant. 

Table~\ref{tab:shakeTest} summarizes the grasps and objects used in the shake test and the results (see also Figure~\ref{fig:GraspRangesFigure}, Right). The grasps were grouped by similarity across participants; different sets had different numbers of grasps within it (Count).

%
%

\subsection{Grasp quality}
In addition to validating the grasps using a shake test we qualitatively assessed grasp quality by directly asking the participants how they felt about the grasp they provided.  Around half of the participants said at least one grasp was not quite what they wanted, particularly for more complex objects such as the plane. The major refrain was that the participants didn't like that the joints in the fingers couldn't be controlled individually (the Barrett fingers are underactuated so the slider bends the entire finger, not each joint independently). This was most noticeable in cases where the finger locks up due to collision --- one part of the finger comes in contact and locks, while the tip or remaining part of the finger stops where it is and doesn't close all the way around the object.  Other issues were the fingers being too thick, the hand too big, or the controls being too fidgety to achieve some of the more precise grasps the participants had intended to perform.

Only one participant commented that they might change their robotic hand grasp after doing the human hand task. We found no evidence for order effects in the data.

\begin{figure}
  \centering
  \includegraphics[width=0.45\linewidth]{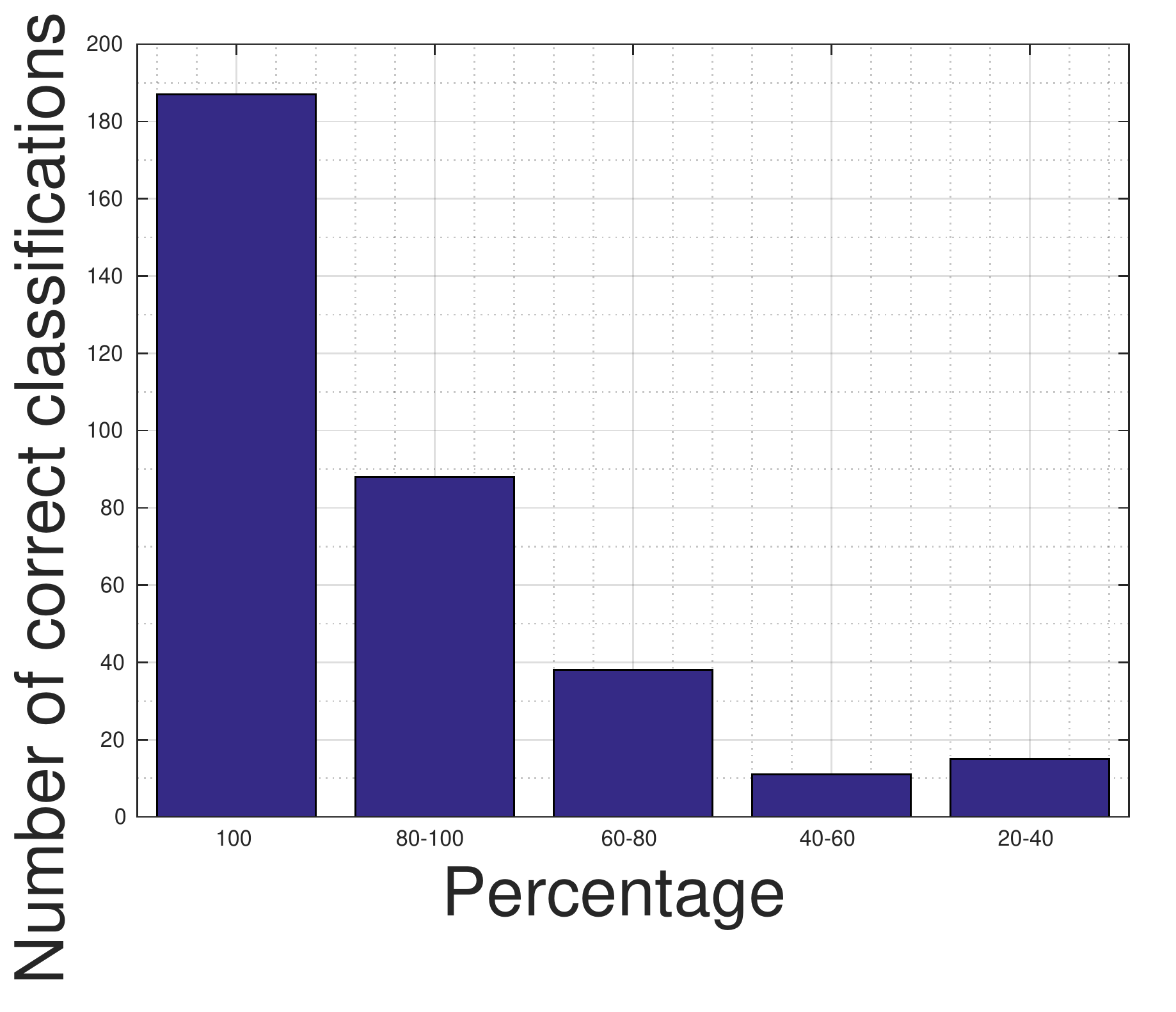}
  \includegraphics[width=0.45\linewidth]{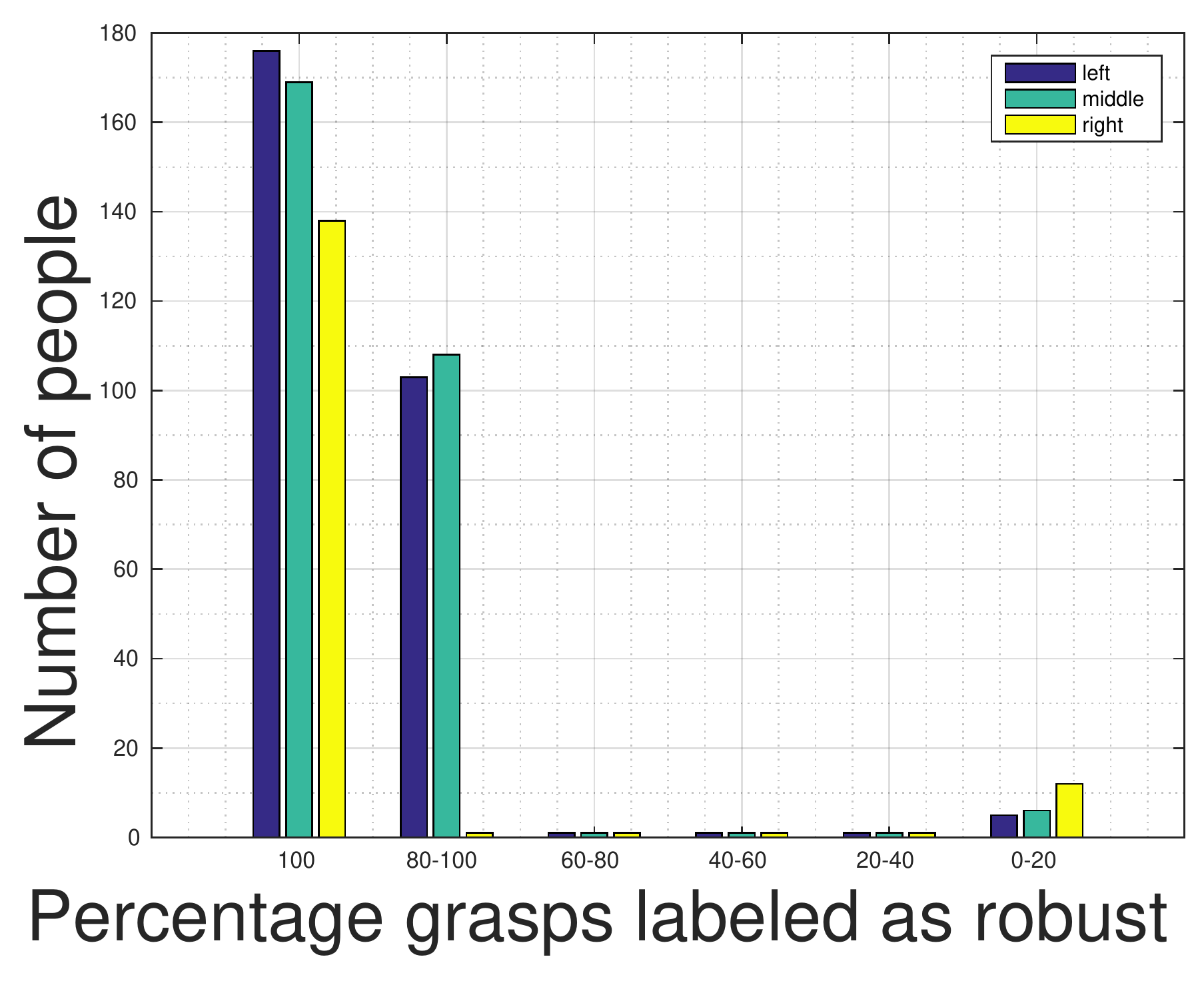}
  \caption{Results of the post capture on-line survey (8-10 answers per grasp). Left: Distribution of intermediate grasps which were correctly classified; over half the grasps had 100\% correct classification. Right: Distribution of grasps that participants said would work. Green is the intermediate grasp; blue and yellow are the optimal and extreme grasps. }
  \label{fig:surveyResults}
  \vspace{-0.1in}
\end{figure}

We also performed a post-capture on-line survey asking participants (not from the study that collected the grasps on the physical robot) to evaluate if the grasps (original, interpolated, and extreme) would work. For this survey, we used OpenRAVE renderings of the hand and object taken from the viewpoint (approximately) of the eye-gaze video. See Figure~\ref{fig:surveyResults}, right.

\begin{table}
\centering
\caption{Shake test results}
\begin{tabular}{|l|cc|ccc|}
\hline
\multicolumn{1}{|>{\centering\arraybackslash}c|}{Object} & Set & Type & Count & Total  & Successful\\
\hline
Spray & 1 & original & 5(2+3) & 25 & 24 \\
Bottle & 2 & original & 2(0+2) & 10 & 10 \\
 & 3 & interpolated & 2(2+0) & 10 & 10 \\
 \hline
Wine & 1 & original & 3(0+3) & 15 & 15 \\
 Glass& 2 & original & 6(3+3) & 30 & 27 \\
 & 3 & original & 3(2+1) & 15 & 15 \\
 & 4 & interpolated & 2(2+0) & 10 &5\\ 
 \hline
Cereal & 1 & original & 6(4+2) & 30 & 30 \\
 Box & 2 & original & 5(3+2) & 25 & 25 \\
& 3&interpolated& 2(2+0) &10 &10 \\
& 4 & interpolated &2(2+0) &10&10 \\
\hline
chunk & 1 & original & 9(6+3) & 45 &  44\\
of & 2 & original & 7(4+3) & 35 & 33 \\
foam & 3 & original & 4(2+2) & 20 & 20 \\
& 4 & interpolated & 2(2+0) & 10 & 10 \\
& 5 & interpolated & 2(2+0) & 10 & 10 \\
\hline
Ball & 1 & original & 7(4+3) & 35 & 35 \\
 & 2 & original & 8(5+3) & 40 & 40 \\
 & 3 & original & 5(3+2) & 25 & 25 \\
 & 4 & original & 5(3+2) & 25 & 21 \\
 & 5& interpolated & 2(2+0)&10&10 \\
 &6&interpolated&2(2+0)&10&10 \\
 \hline
 Total &  &  & 91 & 455 & 434\\
 &&&\multicolumn{2}{c}{Overall grasps:}&(95.38\%) \\
 &&&\multicolumn{2}{c}{Interpolated grasps:}&(93.75\%) \\
\hline
\end{tabular}
\label{tab:shakeTest}
\vspace{-0.1in}
\end{table}


%
%
%
%
%
%

\section{DISCUSSION AND CONCLUSIONS}

\noindent {\bf Protocol variations:}  Randomizing the order of human hand versus robotic one does not seem to produce any noticeable differences in the quality of the data.  We have experimented with doing all human (or all robotic) first; however, this introduces considerable fatigue for the robotic hand phase, so we recommend interleaving them.  Using two (or more) Kinect sensors would also improve the 3D tracking of the object.  


\noindent {\bf Usefulness of data: } Our on-going analysis indicates that there are important differences in the way participants approach placing the robotic hand versus the human one, both in what they gaze at and the types of grasps produced.   Examples of {\em using} the captured data (eg analysis of what people were looking at or thinking, using machine learning on the captured grasp ranges) will be explored in future work.  We will share the captured grasp data using RoboEarth~\cite{Zweigle:2009:RCR:1655925.1655958}.



\noindent {\bf Conclusion:} We have presented a time-efficient protocol for capturing human-specified robot grasps for object-manipulation tasks that captures five different streams of data in the same amount of time. This protocol supports collection of {\em ranges} of grasps, potentially yielding more useful information for machine learning algorithms. The protocol also explicitly aims to capture the human reasoning behind the grasps through three mechanisms: Comparison with human hand grasping, think-aloud protocol, and eye-gaze tracking. We have verified that the extremes of the grasp ranges given by the participants do result in good grasps. 



\section*{APPENDIX: 3D DATA CALIBRATION}

\subsection{Hand to Kinect sensor to object}
\label{sec:PCArmCalibration}

\begin{figure}
  \centering
  \includegraphics[width=0.8\linewidth]{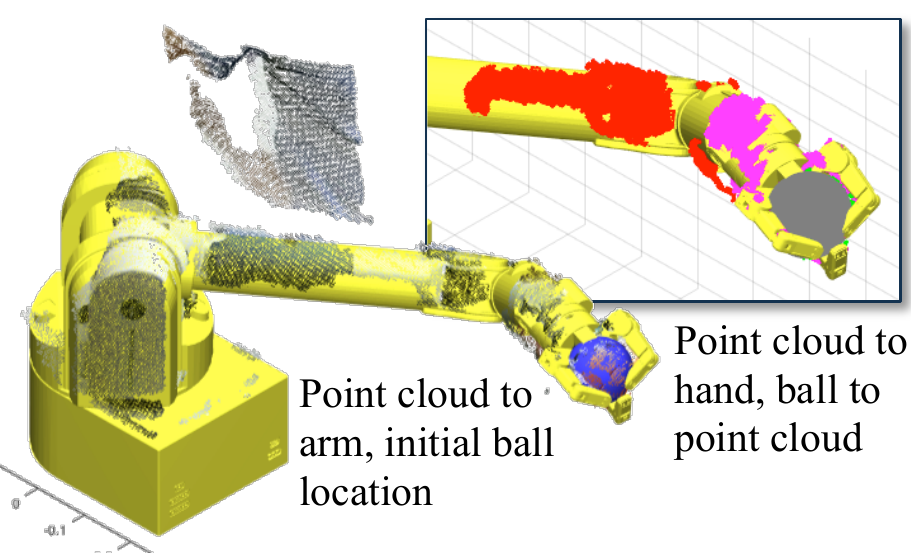}
  \caption{Aligning the Kinect point cloud with the entire arm and hand. Upper right: Aligning the point cloud to the hand and the object to the point cloud by partitioning the point cloud (red arm, purple hand, green object). }
  \label{fig:alignment}
  \vspace{-0.1in}
\end{figure}

Although we know the joint angles of the hand and arm, we do not know where the object is with respect to the hand. We used the Kinect point cloud to calculate an object transform in a three-step process: Generate a polygonal model of the arm from the joint data, align the point cloud to the arm, then align a polygonal model of the object to the point cloud (see Figure~\ref{fig:alignment}). 

%
%
We used OpenRave~\cite{Diankov_2008_6117} to produce a polygonal model of the hand and arm in the position specified by the joint angles. We hand-aligned this model to the Kinect data then used MATLAB's Procrustes alignment to refine this alignment matrix~\footnote{ROS has an alignment function for the Barrett arm, but the sticker was positioned too close to the edge of the image to produce a reliable alignment.} 


We used Iterative Closest Points (ICP)~\cite{icp} to align the arm to the point cloud (trimmed to the arm area). We manually specified a starting transformation for the object then used ICP followed by Procrustes to alternate aligning the point cloud to the hand and the object to the point cloud. 


\addtolength{\textheight}{-12cm}   


\section*{ACKNOWLEDGMENT} Research funded in part by NSF grant CNS 1359480. We would like to thank the participants for their time and effort. We would also like to thank Dr. Reynold Bailey for the generous loan of his SMI glasses, without which this project would not have been possible


\bibliographystyle{IEEEtran}
\bibliography{main,ref1}

\end{document}